\documentclass[a4paper]{jpconf}
\usepackage{graphicx}
\begin{document}

\title{Fully damped Mott oscillations in sub-barrier elastic scattering of identical heavy 
         ions and the nuclear interaction}

\author{M. S. Hussein}

\address{Instituto de F\'{i}sica and Instituto de Estudos Avan\c{c}ados, Universidade de S\~{a}o Paulo, 
S\~{a}o Paulo, Brazil\\
 Departamento de F\'{i}sica, Instituto Tecnol\'{o}gico de Aeron\'{a}utica, CTA, S\~{a}o Jos\'{e} dos Campos, S.P., Brazil}

\author{L. F. Canto}
\address{Instituto de F\'{i}sica, Universidade Federal  do Rio de Janeiro, Rio de Janeiro, Brazil\\
Instituto de F{i}sica, Universidade Federal Fluminense, Av. Gal. Milton Tavares de Souza s/n, Niter\'{o}i, R.J., Brazil
}
\author{W.Mittig}
\address{National Superconducting Cyclotron Laboratory and Department of Physics and Astronomy, Michigan State University, East Lansing, Michigan 48824, USA}

\ead{hussein@if.usp.br}

\begin{abstract}We investigate the possible disappearance of Mott oscillations in the scattering of bosonic nuclei at sub-barrier energies. This effect is universal and happens at a critical value of the Sommerfeld parameter. It is also found that the inclusion of the short-range nuclear interaction has a profound influence on this phenomenon. Thus we suggest that the study of this lack of Mott oscillation, which we call, "transverse isotropy" is a potentially useful mean to study the nuclear interaction.

\end{abstract}

\section{ Introduction}

It is quite well known that the scattering amplitude describing the scattering of particles with a long-range interaction, such as the Coulomb one, composed of  a slowly varying amplitude and a possibly  rapidly varying phase. This results in a quantal cross section which is devoid of oscillations and which exactly coincides with the corresponding classical cross section. The reason for this is understood semi-classically, in the sense that in the evaluation of the quantal amplitude using the stationary phase method, one finds only one dominant contribution. The absence of oscillations in the scattering of non-identical particles with long range interaction, is changed when dealing with identical particles as the scattering amplitude in this case has to be symmetrized (bosons) or anti-symmetrized (fermions). In these cases one encounters rather strong oscillations in the cross section arising from the interference of the amplitude evaluated at an angle $\theta$ and the one at $\pi - \theta$. The resulting, Mott, cross section is oscillatory and can be used to assess the importance of other weaker interactions which may be present. This idea has been explored by several authors in the case of heavy-ion scattering, where the long range interaction is the Coulomb one and the weaker interactions, can be long-ranged, such as multipole polarizability \cite{lynch1982,lynch1982B}, relativistic effects \cite{hussein1984}, color van der Waals force \cite{hussein1990, villari1993}, or short-ranged, such as the nuclear interaction \cite{hussein1984,hussein1984B}. In this contribution we demonstrate that the Mott cross section in the absence of the weaker forces (pure Coulomb) can become structureless at a certain value of the Sommerfeld parameter \cite{canto2001}. This surprising effect can be explored to pin down the properties of the short-range nuclear interaction at energies well below the Coulomb barrier. In the next sections we give the details of our recent investigation \cite{canto2013,canto2014}.

\section{Mott oscillations and their disappearance: Transverse Isotropy}
We consider the scattering of two charged particles (nuclei) with charges $Z_p$ and $Z_t$. The interaction is  predominantly Coulomb. with a small nuclear contribution.  The unsymmetrized scattering amplitude is given by,
\begin{equation}
f_{\rm \scriptscriptstyle{c}}(\theta)=-\frac{a}{2}\  e^{2i\sigma_{0}}\ \
\frac{e^{-i\eta\ln(\sin^{2}\theta/2)}}{\sin^{2}(\theta/2)} = A_{\rm C}(E,\theta) e^{i\Delta_{\rm C} (E,\theta)},
\label{fc}
\end{equation} 
\noindent
where the amplitude, $A_{\rm C}(E, \theta)$ and the phase, $\Delta_{\rm C}(E, \theta)$ are defined in Eq.~(\ref{fc})
\begin{equation}
A_{\rm C}(E, \theta) = - \frac{a}{2\sin^{2}(\theta/2)}
\end{equation}
\begin{equation}
\Delta_{\rm C}(E, \theta) = 2\sigma_{0} -\eta\ln(\sin^{2}\theta/2)
\end{equation}

\noindent
and where $\eta$ and $a$ are respectively the Sommerfeld parameter, and half the distance of closest approach in a head-on collision (zero impact parameter), given by
\begin{equation}
\eta = \frac{Z_p Z_{t}e^2}{\hbar v},\qquad a=\frac{Z_p Z_{t}e^2}{2E_{\rm c.m.}}.
\end{equation}
Above, $v$ is the relative velocity, and $\sigma_{0}$ is the s-wave Coulomb phase shift
\begin{equation}
\sigma_{0}=\arg \big\{ \Gamma(1+i\eta) \big\},
\label{sig0}
\end{equation}

\noindent where $\Gamma$ stands for the usual Gamma-function. 

An excellent approximation for $\sigma_0$ was obtained in \cite{barata2011},

\begin{equation}
\sigma_{0} = \frac{1}{2}\tan^{-1}\left( \eta \right)
+ \eta\bigg( \ln\left(\sqrt{ 1 + \eta^2 }\right) -1\bigg)
- \frac{\eta} {12\,\left(1+\eta^2\right)} .
\label{sig1}
\end{equation}

The Coulomb amplitude, Eq.~(\ref{fc}), must be symmetrized in the case of the scattering of identical bosonic charged particles, which for spin zero, $f(\theta) = f_{\rm C}(\theta) + f_{\rm C}(\pi - \theta)$. The resulting Mott cross section, $\sigma_{\rm M} (\theta)$ contains an incoherent, classical, piece, $\sigma_{\rm inc}(\theta)$, plus an interference one, $\sigma_{\rm int}(\theta)$, viz,

\begin{equation}
\sigma_{\rm M}(\theta) = \sigma_{\rm inc}(\theta) + \sigma_{\rm int}(\theta),
\label{inc-int}
\end{equation}
where $\sigma_{\rm inc}(\theta)$ is the incoherent sum of contributions from the two amplitudes, $f_{\rm C}(\theta)$, and $f_{\rm C}(180^{\rm o} - \theta)$,
\begin{equation}
\sigma_{\rm inc}(\theta) = \big| f_{\rm C}(\theta)\big|^2+\big|f_{\rm C}\left( 180^{\rm o}-\theta \right)  \big|^2,
\label{sig_inc}
\end{equation}
and $\sigma_{\rm int}(\theta)$ is the interference term,
\begin{equation}
\sigma_{\rm int}(\theta) = 2\,{\rm Re}\Big\{ f_{\rm C}^\ast (\theta)\,\times f_{\rm C} \left(180^{\rm o}-\theta \right) \Big\}.
\end{equation}
Note that the incoherent part of the cross section is positive-definite, whereas the interference term
may assume positive or negative values.\\

\begin{figure}[th]
\centering
\includegraphics[width=12 cm]{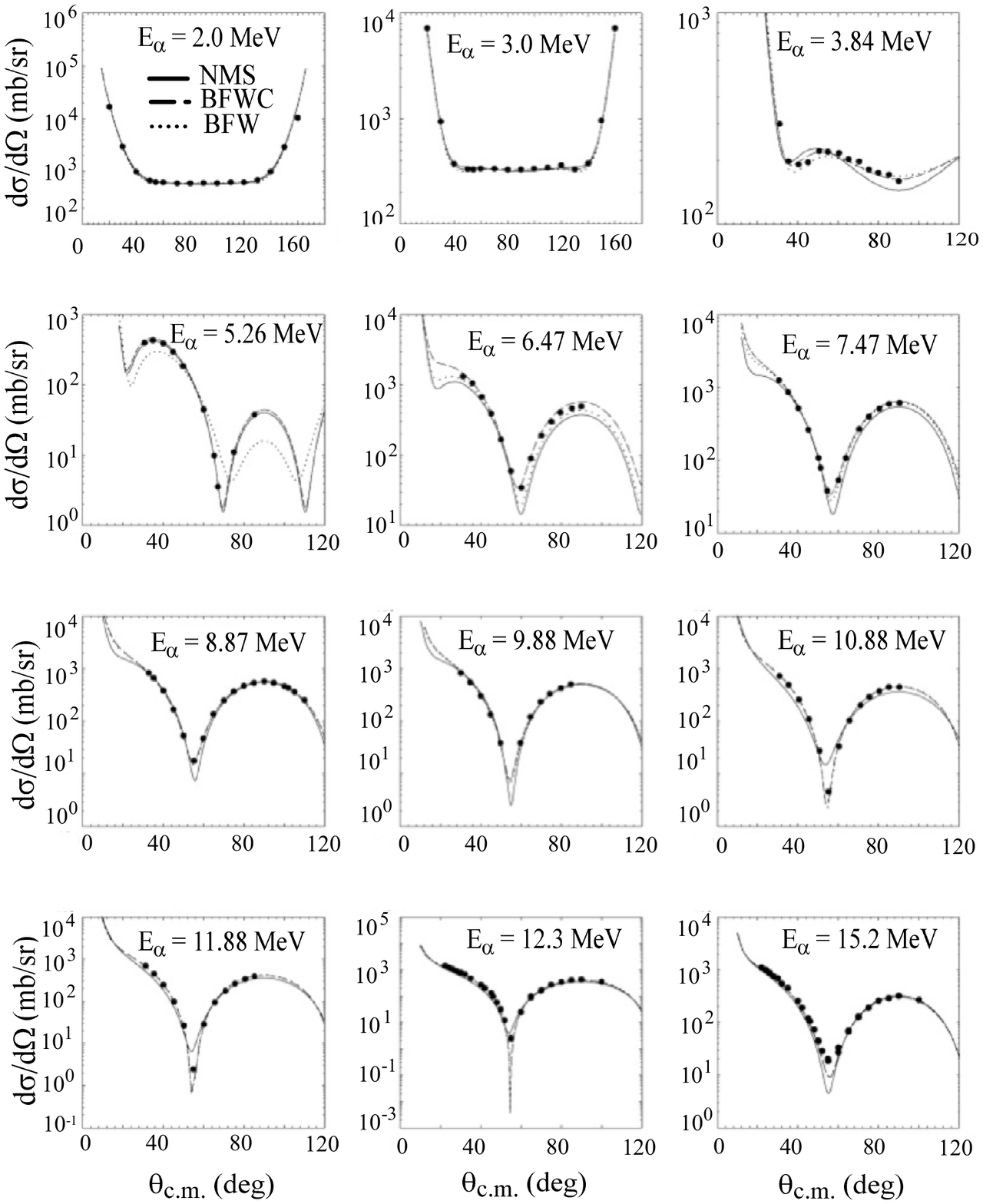}
\caption{(Color online) The experimental angular distributions for the $^4$He+$^4$He system at several
 collision energies. The figure was taken from Abdullah {\it et al.}~\cite{abdullah2006}. For details see the text.}
\label{abdullah}
\end{figure}

It was found in Ref. \cite{canto2001} that the full Mott cross section becomes flat at a critical value, $\eta_c = \sqrt{2}$, of the Sommerfeld parameter. This feature was coined $\textit{Transverse Isotropy}$, TI.This was accomplished by demanding that the second derivative of the cross section with respect to $\theta$ to be zero at $\theta = 90^{\rm o}$. Of all the heavy-ion systems which were considered, the best case for studying the TI was found to be the elastic scattering of two $^{4}$He nuclei, for which the condition $\eta_c =\sqrt{2}$ implies a center of mass energy of $E_{\rm c.m.} $= 0.397 MeV, well below the height of the Coulomb barrier, $V_{ B}$ = 0.87 MeV. This guarantees that, for all practical purposes, the scattering is purely Coulomb. Our calculation also indicated that at an energy below $E_c$ the cross section at $90^{\rm o}$ exhibits a maximum, while above the critical energy it exhibits a minimum. In 2006, Abdullah et al.  \cite{abdullah2006} published elastic data for the $\alpha + \alpha$ system at several energies starting from around the Coulomb barrier. In Fig.~(\ref{abdullah}) we show their results,

We see two flat regions followed by the usual oscillatory behavior. The first flat region is at $E_{\rm c.m.}$ = 1.0 MeV, and the second at $E_{\rm c.m.}$ = 1.5 MeV. Our calculation for a pure Coulomb potential Mott scattering predicts one flat region at $E_{\rm c.m.} \approx$ 0.4 MeV. This clearly indicates the influence of the short range nuclear interaction.

\section{Inclusion of the nuclear interaction}
In this section we investigate the influence of the nuclear interaction on the TI. We do this by adding this potential to the Coulomb one and solve the scattering problem. From the analytical point of view the determination of regions of the TI is more difficult here, but the calculation of the second derivative can be easily done numerically. First we give the salient feature of the symmetrized cross section in the presence of the nuclear interaction.\\

Using as a guide the form of the Rutherford cross section, Eq. (\ref{fc}), we can write the symmetrized cross section as 

\begin{equation}
\frac{d\sigma}{d\Omega} = \sigma_{\rm inc} + \sigma_{int}
\end{equation}
\noindent
with the incoherent part $\sigma_{\rm inc}$, affected only very little by the short range nuclear interaction at sub-barrier energies,

\begin{equation}
\sigma_{\rm inc} = \frac{a^2}{4}\left[\frac{1}{\sin^4{(\theta/2)}} + \frac{1}{\cos^4{(\theta/2)}}\right]
\end{equation}
\noindent
and the coherent, interference, part $\sigma_{\rm int}$,
\begin{equation}
\sigma_{\rm int} = 2\sqrt{\sigma_{\rm C}(\theta)\sigma_{\rm C}(180^{\rm o} -\theta)}\cos{\Bigg([\Delta_{\rm C}(\theta) + \Delta_{\rm N}(\theta)] -  [\Delta_{\rm C}(180^{\rm o} - \theta) + \Delta_{\rm N}(180^{\rm o} - \theta)] \Bigg)}
\end{equation}
\noindent
which gives, with the help of Eq.~(\ref{fc}), 
\begin{equation}
\sigma_{inc}(\theta) =  \frac{a^2}{2}\frac{1}{\sin^2{(\theta/2)}\cos^2{(\theta/2)}} \cos{\left(2\eta\ln\tan\frac{\theta}{2} + [\Delta_{\rm N}(\theta) - \Delta_{\rm N}(180^{\rm o} - \theta)]\right)}
\end{equation}
The second derivative of the full cross section at $\theta = 90^{\rm o}$ is evaluated for a nuclear interaction used in Ref.~\cite{abdullah2006}. They employed the following nuclear force,
\begin{equation}
 V_{\rm N}(r) = - V_A e^{-(r/R_A)^2} + V_R e^{-(r/R_R)^2} 
 \label{nucl-pot}
 \end{equation}
\noindent 
where, $V_A$ = 122.62 MeV, $R_A$ = 2.132 fm, $V_R$ = 3.0 MeV, and $R_R$ = 2.0 fm, and a Coulomb interaction behaving as $4e^2/r$ for $r > R_C$, with $R_C$ = 5.8 fm and for $r < R_C$,
 
 \begin{equation}
 V_C(r) = \frac{4e^2}{2R_C}\left[3 - \left(\frac{r}{R_C}\right)^2\right]
 \label{could-pot2}
 \end{equation}
 
\begin{figure}[th]
\centering
\includegraphics[width=8 cm]{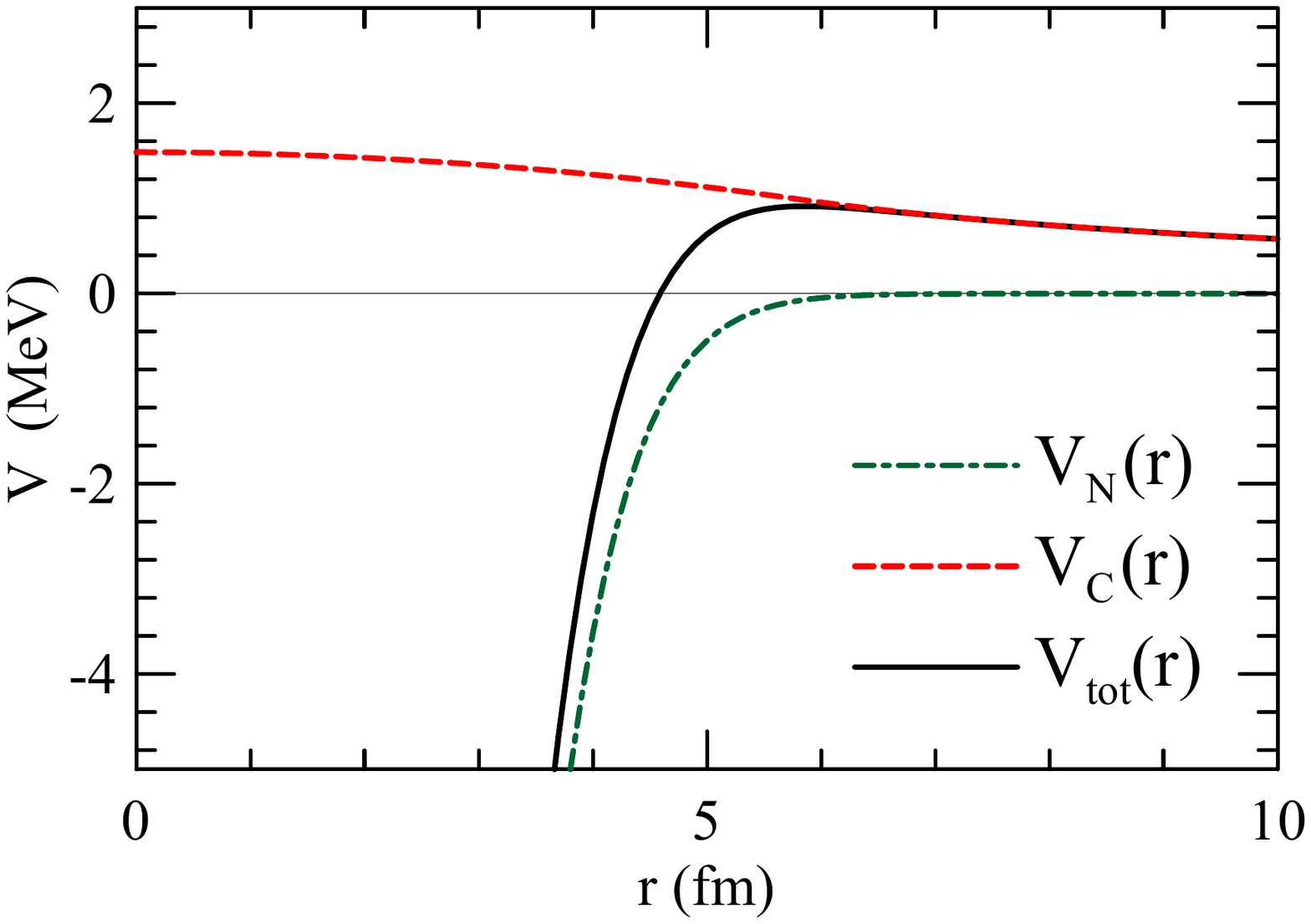}
\caption{(Color online) The $\alpha + \alpha$ potential of Ref. \cite{abdullah2006}. The nuclear potential is the sum of two Gaussians, Eq.(\ref{nucl-pot}). The Coulomb potential is taken as $V_{\rm C}(r) = \frac{4e^2}{r}\Theta(r -R_C) + \frac{4e^2}{2R_C}[3 - (r/R_C)^2]\Theta(R_C - r)$, where $\Theta(x)$ is the step function.}
\label{potentialAbdullah}
\end{figure}

\noindent 
The above potentials, $V_{\rm N}(r)$, $V_{\rm C}(r)$, and $V_{\rm N}(r) + V_{\rm C}(r)$ are shown in Fig. (\ref{potentialAbdullah}).

The nuclear phase is directly related to the nuclear potential and, to first order, is given by,
 
 \begin{equation}
 \Delta_{\rm N} = - \sqrt{\frac{2\mu}{\hbar^2}}\int_{r_0}^{\infty}\frac{V_{\rm N}(r)}{\sqrt{E - V_{\rm C}(r) - \frac{\hbar^2 l(l + 1)}{2\mu r^2}}}dr
\end{equation} 
\noindent
where $r_0 = 2a$ is the distance of closest approach for head-on collision.

The above formulae can be used to assess the importance of the nuclear interaction on the Mott scattering. In fact, this procedure, based on perturbation theory, was used in Refs.~ \cite{hussein1990, villari1993,hussein1984,hussein1984B}. It would be instructive to perform the calculation of the second derivative of the cross section and demands that it be zero at $\theta$ = 0.0, to find the nuclear modified critical Sommerfeld parameter, which would be $\eta_c = \sqrt{2} + \delta\eta(E)$. We leave this exercise for a future work. In the following, however, we perform the calculation exactly, by solving the scattering Schr\"{o}dinger equation (optical model calculation)

\begin{figure}[th]
\centering
\includegraphics[width=8 cm]{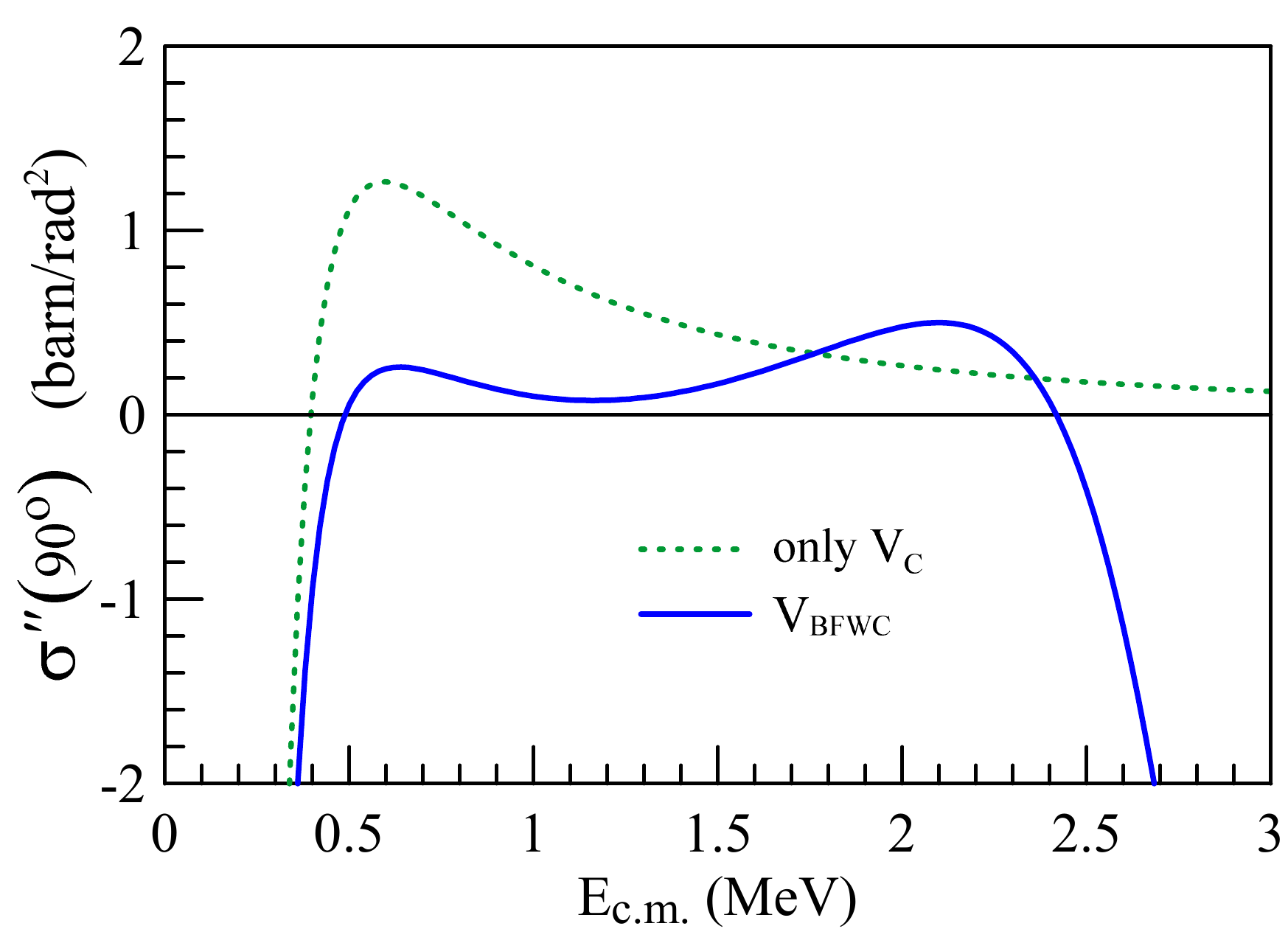}
\caption{(Color online) Second derivative of the Mott cross section at $\theta = 90^{\rm o}$ for the BFWC potential of Fig. (\ref{potentialAbdullah}). The dotted line is the usual Mott (only Coulomb) result, while the solid line takes
into account both the Coulomb and the Nuclear potentials.}
\label{dersecGauss}
\end{figure}

The result of the optical model calculation of the second derivative of the symmetrized cross section at $\theta = 90^{\rm o}$ is given in Fig.(\ref{dersecGauss}). We see two clear crossings of the horizontal axis at 0.5 MeV and at 2.5 MeV. These crossing points correspond to TI . Between these two points, at $E_{\rm c.m.} \approx$ 1.4 MeV, the second derivative attains a very small value, very close to zero. We may take this point as a third TI. It is interesting to note that the TI at the lowest energy, $E_{\rm c.m.}$ = 0.4 MeV, is shifted by the nuclear interaction to $E_{\rm c.m.}$ = 0.5 MeV, indicating a slightly smaller value of $\eta_c$ than $\sqrt{2}$. The data of Ref.~\cite{abdullah2006} exhibit two TI's, one at $E_{\rm c.m.}$ = 1.0 MeV, and the other one at 1.5 MeV. Had the authors of Ref.~\cite{abdullah2006} extended their measurement to energies, below the barrier, they would probably have encountered the TI at around (0.4 - 0.5) MeV. It would therefore  be quite interesting to perform measurements of the elastic scattering of $\alpha + \alpha$ at around $E_{\rm c.m.} \approx$ 0.4 MeV and verify our findings.\\

\section{Conclusions}
In this work, we summarized our research on the phenomenon of Transverse Isotropy, predicted to occur in the low energy scattering of identical charged particles. At a critical value of the Sommerfeld parameter $\eta = \sqrt{2}$, the cross section becomes quite flat. The c.m. energy corresponding to this value of $\eta$, is found to be below the height of Coulomb barrier only for a very few systems. We considered the system $\alpha + \alpha$ at sub-barrier energies and compared our findings to the experimental data of \cite{abdullah2006}. We found that whereas our calculation exhibits only one flat region, the data, taken at energies above the barrier, show two flat regions. The region we have predicted at $E_{\rm c.m.}$ = 0.4 MeV was not accessed by the experimental group. The difference in the results of our calculation and the data is attributed to the importance of the nuclear interaction even sub-barrier energies. By including the nuclear interaction in our calculation, we accounted reasonably well for several features of the data. This clearly indicates that the study of TI could shed light on the nuclear interaction at very low energies. \\

\section*{Acknowledgments}
This work was supported by the CNPq, FAPESP, CAPES/ITA, FAPERJ and UFF.

\end{document}